\documentclass[twocolumn,prb,amsmath,amssymb,floatfix,superscriptaddress]{revtex4-1}
\usepackage{color}
\usepackage{graphicx}
\usepackage{dcolumn}
\usepackage{bm}
\usepackage{times}
\begin{document}

\title{Thermal evolution of the full three-dimensional magnetic
excitations in the multiferroic BiFeO$_3$}

\author{Zhijun~Xu}
\affiliation{Condensed Matter Physics and Materials Science
Department, Brookhaven National Laboratory, Upton, New York 11973,
USA}
\author{Jinsheng~Wen}
\affiliation{Physics Department, University of California, Berkeley, CA 94720, USA}
\affiliation{Materials Science Division, Lawrence Berkeley National 
Laboratory, Berkeley, CA, 94720, USA}
\author{Tom Berlijn}
\affiliation{Condensed Matter Physics and Materials Science
Department, Brookhaven National Laboratory, Upton, New York 11973,
USA}
\author{Peter M. Gehring}
\author{Christopher Stock}
\affiliation{NIST Center for Neutron Research, National Institute of
Standards and Technology, Gaithersburg, Maryland 20899, USA}
\author{M. B. Stone}
\affiliation{Quantum Condensed Matter Division, Oak Ridge National Laboratory,
Oak Ridge, TN, 37831, USA}
\author{Wei Ku}
\author{Genda~Gu}
\author{Stephen M. Shapiro}
\affiliation{Condensed Matter Physics and Materials Science
Department, Brookhaven National Laboratory, Upton, New York 11973,
USA}
\author{R. J. Birgeneau}
\affiliation{Physics Department, University of California, Berkeley, CA 94720, USA}
\affiliation{Materials Science Division, Lawrence Berkeley National 
Laboratory, Berkeley, CA, 94720, USA}
\author{Guangyong~Xu}
\affiliation{Condensed Matter Physics and Materials Science
Department, Brookhaven National Laboratory, Upton, New York 11973,
USA}
\date{\today}

\begin{abstract}

\end{abstract}

\maketitle

{\bf The idea of embedding and transmitting information within the
fluctuations of the magnetic moments of spins (spin waves) has been recently
proposed~\cite{Magnonics} and experimentally tested.~\cite{Kajiwara}  The
coherence of spin waves, which describes how well defined these excitations
are, is of course vital to this process, and the most significant factor that
affects the spin-wave coherence is temperature.  Here we present neutron
inelastic scattering measurements of the full three-dimensional spin-wave
dispersion in BiFeO$_3$, which is one of the most promising functional
multiferroic material~\cite{Catalan},  for temperatures from 5\,K to 700\,K.
Despite the presence of strong electromagnetic coupling,  the magnetic 
excitations behave like conventional magnons over all parts of the Brillouin 
zone. At low temperature the spin-waves are well-defined coherent modes, 
described by a classical model for a G-type antiferromagnet. 
A spin-wave velocity softening is already present at
room temperature, and more pronounced damping occurs as the magnetic ordering
temperature T$_N \sim 640 $\,K is approached.  In addition, a strong
hybridization of the Fe $3d$ and O $2p$ states is found to modify the
distribution of the spin-wave spectral weight significantly, which implies
that the spins are not restricted to the Fe atomic sites as previously
believed.}

Coupling between static magnetic and ferroelectric orders in multiferroic
materials has attracted tremendous recent interest because of its potential
use in device applications.~\cite{datastorage,Eerenstein,swcheong}  If such
coupling were to extend to dynamical properties, e.\ g.\ between lattice
vibrations and spin fluctuations, then a new type of excitation may emerge.
Such hybrid excitations, or ``electromagnons,'' have already been studied in a
number of multiferroic systems using dielectric and Raman
measurements.~\cite{Pimenov,Cazayous,Kumar1,Rovillain}  An even more
fascinating aspect of this novel concept is the potential ability to control
the spin waves using the electric fields generated by low-power dissipating
electronic circuits instead of magnetic fields, a possibility that has been
suggested for multiferroic BiFeO$_3$ at room
temperature,~\cite{Rovillain,BiFeO3_film_field} making this material
of particular interest~\cite{Catalan}. Unfortunately optical
measurements such as Raman scattering are only sensitive to long-wavelength
($q\sim 0$) excitations. Neutron scattering techniques, on the other hand,
provides a powerful tool to probe these hybrid excitations over a large 
momentum space. There have already been some pioneering neutron scattering
measurements on the magnetic excitations from powder BiFeO$_3$ 
samples~\cite{Delaire},
as well as single crsytal samples~\cite{cheong,Matsuda} but measured only around 
room temperature. 
A more complete picture of these modified spin-wave
excitations, i.\ e.\ the electromagnons, at all wave vectors ${\bf Q}$, and
especially their evolution with temperature is highly desired.  

\begin{figure*}[t]
\includegraphics[width=\linewidth]{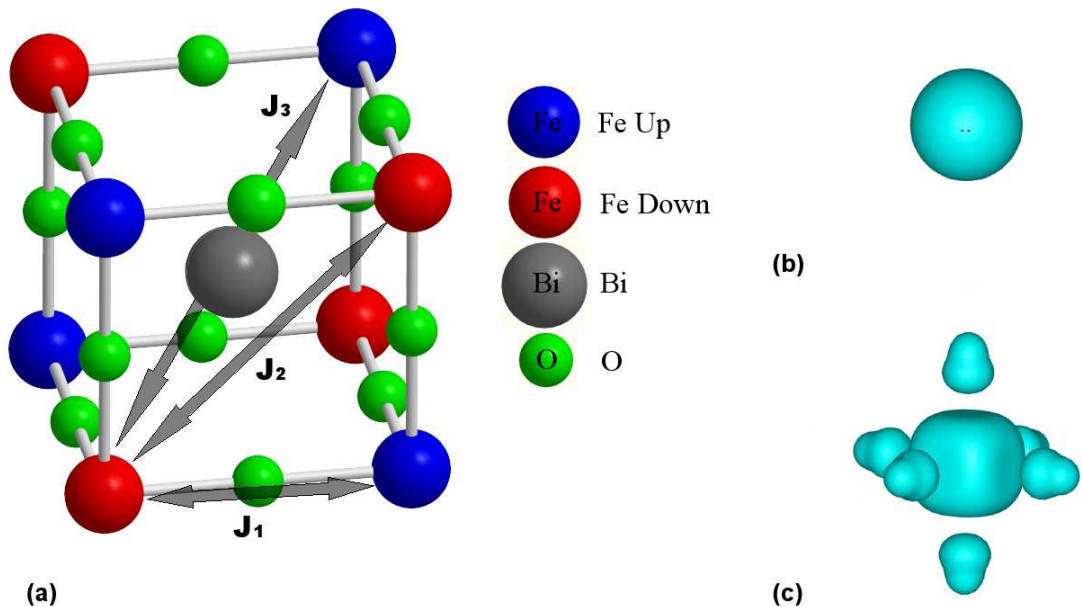}
\caption{(Color online) Schematics of the BiFeO$_3$ crystal structure (a), and 
spin densities from conventional Fe ionic model (b) and from 
first principles Wannier analysis  taking
into account the O {\it 2p} orbital (c).} \label{fig:1}
\end{figure*}

BiFeO$_3$ becomes ferroelectric at T$_C \sim 1100$\,K. Below T$_C$ this
material exhibits an $R3C$~\cite{BFO_Structure1,BFO_Structure2} 
crystal structure, which can be
viewed as a (weakly) rhombohedrally-distorted perovskite structure in which
the oxygen octahedra are also distorted.  For the purpose of our magnetic
analysis this is well approximated by the pseudo-cubic unit cell shown in
Fig.~\ref{fig:1} (a). The N\'{e}el temperature for BiFeO$_3$ is T$_N \sim
640$\,K, and the ordered magnetic phase has a cycloid
modulation~\cite{spiral} with a long periodicity that is superimposed on a 
simple G-type antiferromagnetic structure. The magnetic Bragg peaks appear 
close to the magnetic zone-centers of ${\bf Q_{AF}}=(0.5,0.5,0.5)$~(r.l.u.).

\begin{figure}[ht]
\includegraphics[width=\linewidth,trim=-2cm -2cm 0cm -1cm,clip]{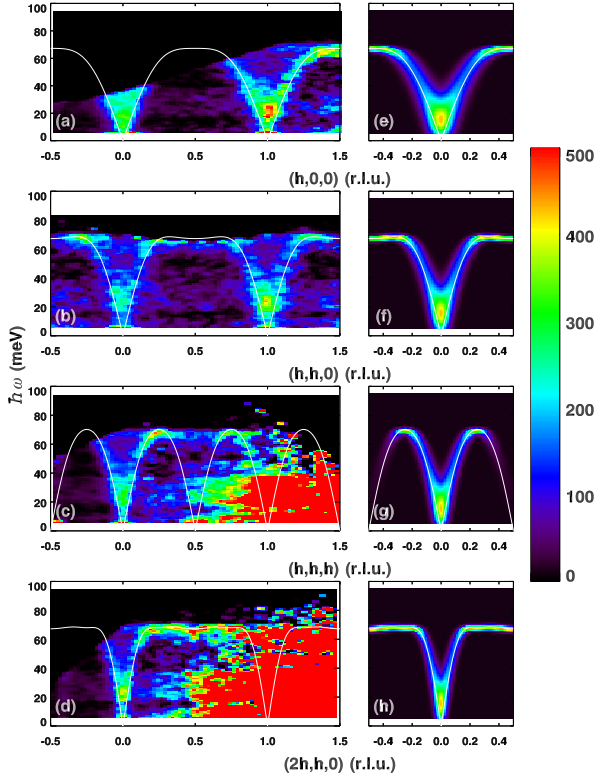}
\caption{(Color online) Magnetic excitations from BiFeO$_3$ at T=5~K. 
The vertical axes denote energy transfer, while the horizontal 
axes are ${\bf Q} - {\bf Q_{AF}}$ for ${\bf Q}$ along [100], [110], [111], and [210] directions for (a), 
(b), (c) and (d), respectively. The strong background appearing at large 
${\bf Q}$ comes from phonon scattering. (e) to (h) are model calculations 
described in the text.   } \label{fig:2}
\end{figure}

In Fig.~\ref{fig:2} (a)-(d), the magnetic excitations measured at $T=5$~K
along [001], [110], [111], and [210] directions are plotted. At base temperature,
the spin-wave
excitations are well defined, and disperse quickly when moving 
away from ${\bf Q_{AF}}$.  One will see that although the features at higher
energies near the top of the bands are very well resolved, fine features
at low energies for ${\bf Q}$ near ${\bf Q_{AF}}$ cannot be accurately 
determined, due to our relatively coarse energy and wave-vector resolution.
We then choose to model the magnetic Hamiltonian by 
ignoring the effect of the long-period cycloid 
modulation, or any single-ion spin anisotropy in this system, which would 
only affect~\cite{cheong,Matsuda} the magnetic excitations at low energies and in
a narrow range around ${\bf Q_{AF}}\pm\delta$, where $\delta \sim 0.004$ along 
$\langle 110\rangle$ are the modulation wave-vectors for the cycloid structure\cite{Ratcliff1}. 
The Heisenberg spin Hamiltonian of the system therefore can
be simplified to that of a classical G-type AFM,
$H=J\sum_{n,m}{\bf S_n}\cdot{\bf S_m}$, where we only consider the nearest 
neighbor interaction $J_1$, second
nearest neighbor interaction $J_2$, and the third nearest neighbor interaction
$J_3$, as shown in Fig.~\ref{fig:1} (a). The
interaction that dominates is clearly the nearest-neighbor
exchange interaction $J_1$ between Fe spins separated along the
$\langle100\rangle$ direction, whose magnitude can be roughly estimated 
from the top of the magnetic dispersion band, while $J_2$ and $J_3$ can 
also be obtained if the full dispersion is known (see Methods section for 
details). 
We have performed an overall fitting to our
data using the model described in the  Methods section to determine the 
exchange constants.
The model calculations are shown in Fig.~\ref{fig:2} (e)-(h), with 
fitting parameters shown in Table~\ref{tab:1}.
The calculated dispersion curves are also plotted in Fig.~\ref{fig:2}
as solid lines. It is apparent that our model provides a good description
to the data at 5~K. The fact that $J_1$, $J_2$ and $J_3$ are all positive, 
suggests that there is frustration in the system.  Specifically, the 
face-diagonal Fe-Fe
spins are aligned along the same direction in spite of the fact that the
interactions between them are antiferromagnetic.

\begin{figure}[hbt]
\includegraphics[width=\linewidth,trim=-2cm -2cm 0cm -1cm,clip]{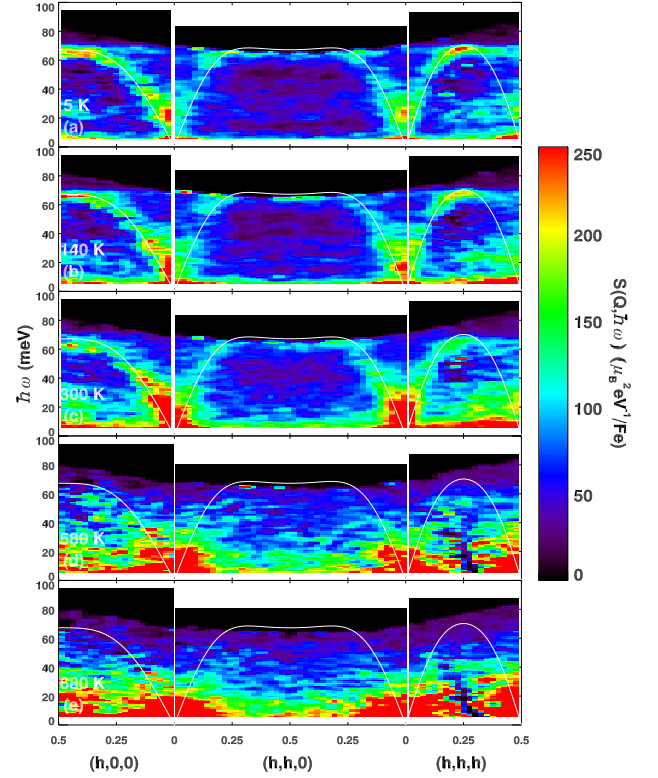}
\caption{(Color online) Magnetic excitations from BiFeO$_3$ measured at
5~K, 140~K, 300~K, 580~K, and 680~K. The horizontal axis denotes
${\bf Q} - {\bf Q_{AF}}$. The solid lines are dispersion calculated
based on parameters obtained from the base temperature (5~K) data.
 } \label{fig:3}
\end{figure}

Upon heating the spin-wave is modified gradually. The intensity
of low energy magnetic excitations is boosted by the Bose factor from 5~K to
140~K and then 300~K, while the overall shape still remain unchanged [see
Fig.~\ref{fig:3} (a) - (c)]. This is 
expected and can be understood as intensities from the magnetic Bragg
peak are shifted into the inelastic channels when the static order 
becomes less robust with heating. At higher energies near the band top, the 
change is minimal between 5~K and 140~K, but at 300~K the band top
is noticeably lower [Fig.~\ref{fig:3} (c)], and the excitations 
become broader in energy. The softening of the spin-wave
modes is due to the renormalization that occurs at non-zero temperatures, which
is essentially a result of interactions between the excited magnons, and
can be modeled with various theories. Instead of going into
details of the renormalization, we would still use the same simple spin-wave
model that describes the low-T data well, but with a set of renormalized 
effective exchange parameters shown in Table~\ref{tab:1}. At 140~K
the effective exchange parameters are virtually the same as those obtained 
at 5~K. The effective exchange constants $J_1$ and $J_2$ both show a clear
drop at T=300~K, and a broadening in energy of $\Gamma \sim 5$~meV is required
to obtain a good fit to the data. Our results indicate that 
although the spin-waves are still relatively well defined at room temperature,
the lifetime (inverse of the energy width) of the spin-wave mode does decrease
significantly compared to that at base 
temperature. The magnitude of the spin-wave softening is smaller than the 
softening observed of the electromagnon modes~\cite{Cazayous} by Raman measurements 
near $q=0$, but note that the energy of the $q=0$ mode in the Raman work
is also related to the cycloid wave-vector~\cite{sousa:012406} $\delta$, 
which also drops with heating~\cite{Ratcliff1}. 
The clear spin-wave velocity softening at 300~K makes the low temperature 
measurements important to obtain the ground state values of the exchange 
constants and related parameters.

With further heating towards $T_N\sim 640$~K, the spin-waves become heavily 
damped, as shown in Fig.~\ref{fig:3} (d) and (e). The data taken at 580~K
already show that although the scattered intensities
are more intense around $\bf Q_{AF}$ due to the dominating 
AFM correlations, well defined spin-waves are no longer present. 
While the N\'{e}el temperature marks the disappearance of 
static magnetic order in the system, the coherence of spin-wave excitations 
already diminishes at temperatures well below $T_N$. 

\begin{figure}[htb]
\includegraphics[width=\linewidth]{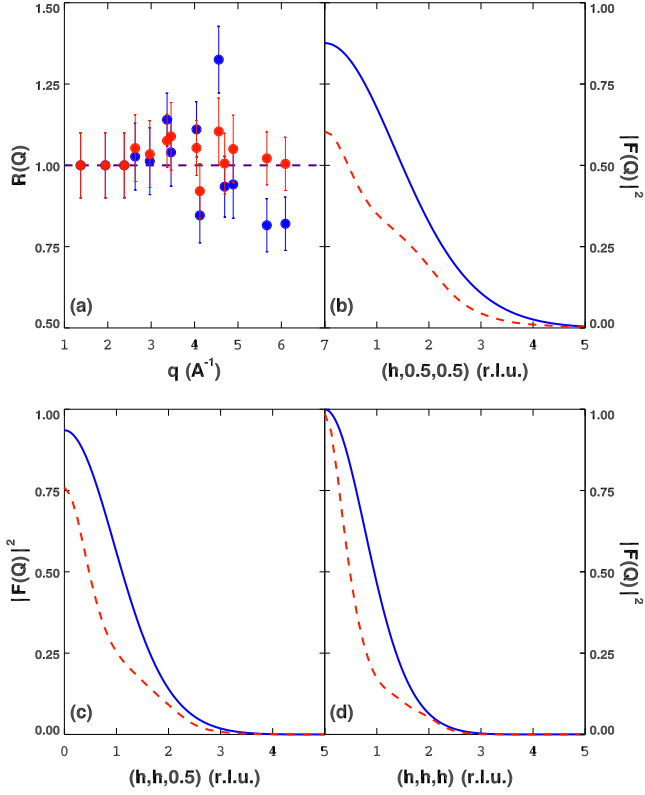}
\caption{(Color online) Comparison of the original ionic Fe form factor and
the new Wannier form factor. (a) Plotting of the ratio $R({\bf Q})=
\frac{I({\bf Q})/|F({\bf Q})|^2}{I({\bf Q_0})/|F({\bf Q_0})|^2}$, 
where $I({\bf Q})$ is magnetic 
scattering intensity measured at wave-vector
${\bf Q}$, while ${\bf Q_0}$ is the wave-vector equivalent to ${\bf Q}$ 
in the first Brillouin zone. Ideally this value should be 1 (see the dashed line).
The red symbols are values calculated using the new Wannier form factor
while the blue symbols using the old Fe ionic form factor. (b) to (d): 
$|F({\bf Q})|^2$ plotted along different directions. } \label{fig:4}
\end{figure}

In addition to probing the coherence of the spin-wave excitations, which are 
collective magnetic responses from the system, our neutron scattering 
measurements also provide information on the local spin structure around 
each Fe atom.
The magnetic form factor $|F({\bf Q})|^2$ describes the Fourier transformation
of the   spatial distribution of spin (density)
around each magnetic ion (in our case, Fe$^{3+}$), into momentum space.
In Fig.~\ref{fig:4} (b-d), we show the magnetic form factor for 
a Fe$^{3+}$ ion with the blue solid lines. This magnetic form factor is 
isotropic and only depends on the length of ${\bf Q}$. However, recent 
analysis of inelastic neutron scattering data from high-T$_C$ cuprates 
suggests~\cite{Igor} that the spin distribution around magnetic ions in metal
oxides can be greatly affected by the presence of the oxygen ion, and could
result in different effective magnetic form factors.

\begin{table}[ht]
\caption{Parameters obtained from modeling the data at various temperatures.
Note that in our spin Hamiltonian, 
one pair of spins ${\bf S_n}$ and ${\bf S_m}$ are counted twice, and their 
binding energy
becomes $H_{nm}=J_{nm}({\bf S_n}\cdot{\bf S_m}+{\bf S_m}\cdot{\bf S_n})=
2J_{nm}{\bf S_n}\cdot{\bf S_m}$, so that the exchange constant $J$ used 
here have same definition as those used in Ref.~\onlinecite{Delaire}, 
but should be compared to half of the $J$ values used in 
Ref.~\onlinecite{cheong}. The units are all meV except for 
the ordered moment $\langle M_z\rangle$, 
which is shown as multiples of $\mu_B$. }
\begin{ruledtabular}
\begin{tabular}{cccccccc}
  & 5~K &140~K & 300~K &580~K & 680~K \\
\hline
$\langle M_z\rangle$& 3.9(2) & 3.6 (2) & 3.2 (3) & 1.9 (4)&-\\
$J_1$ & 2.17(9) & 2.17(9) & 1.9(1) & 1.5(2) & -\\
$J_2$ & 0.11(3) & 0.11(3) & 0.05(5) & 0.03(9) & -\\
$J_3$ & 0.29(9) & 0.29(9) & 0.3(2) & 0.2(3) & -\\
$2J_1\langle S_0\cdot S_1\rangle$&-21(1)&-21(1)&-18(1)&-12(2)&-\\
$2J_2\langle S_0\cdot S_2\rangle$&0.26(8) & 0.26(8)& 0.1(1) &0.1(2)&-\\
$2J_3\langle S_0\cdot S_3\rangle$&-2.8(8)&-2.8(8)&-3(1)&-2(2)&-\\
$\Gamma$&0 &0 &5 (2)&10(5) & -\\
\end{tabular}
\end{ruledtabular}
\label{tab:1}
\end{table}

To compute the magnetic form-factor that includes the effects of the 
crystalline environment, a first principles Wannier analysis~\cite{WeiKu,Igor} 
is employed. 
From LDA+U calculations of BiFeO$_3$~\cite{LDA1,LDA2} the Fe is found to be in the 
high-spin $S=5/2$ state. The magnetic form-factor is obtained by 
Fourier transforming the spin density of the low-energy Wannier functions 
of Fe. In Fig.~\ref{fig:1} (b) and (c), we compare the atomic spin-density 
with the Wannier function derived spin-density. 
The main difference is that the Wannier function derived spin-density spreads 
into the neighboring oxygen atoms, which suppresses the magnetic form-factor 
at none-zero momenta. Direct comparison of the form factors are plotted in 
Fig.~\ref{fig:4} (b)-(d). We can see that for most $Q$, the Wannier 
form factor is 
smaller than the previously used ionic form factor. An experimental verification
of the new form factor is shown in Fig.~\ref{fig:4} (a). Here we plot the ratio
$R({\bf Q})=\frac{I({\bf Q})/|F({\bf Q})|^2}{I({\bf Q_0})/|F({\bf Q_0})|^2}$, 
where 
$I({\bf Q})$ is magnetic scattering intensity measured at wave-vector
${\bf Q}$, while ${\bf Q_0}$ is the wave-vector equivalent to ${\bf Q}$ 
in the first Brillouin zone. It is also worth noting that this type of calculation
is model (Hamiltonian) independent, and $R({\bf Q})$ is supposed to always be
$1$ if the magnetic form factor used is correct. The 
results from the old ionic (blue symbols) and new Wannier (red symbols)
magnetic form factors clearly suggest that the new form factor provides 
a better description of the ${\bf Q}$ dependence.

Using this hybrid magnetic form factor, the ordered moment per Fe at various
temperatures is calculated and listed in the first row of Table~\ref{tab:1}. At T=5~K, this 
value $M_z=4~\mu_B$ corresponds to $S_Z=2.0$, consistent with
the moment obtained from other measurements~\cite{Fischer,Sosnowska}, 
and close to the theoretical 
limit of $S_Z=S=5/2$, using a $g-$factor of $2$. 
Frustration due to a positive (AFM) $J_2$ is likely the main 
reason for the difference. We can also obtain an integrated spectral weight 
from all the spin-wave excitations, to be $S_{dynamic} \sim 2.5$ (i.e. 
$M_{dynamic}^2=S_{dynamic}g^2\mu_B^2=10\mu_B^2$). Combining the two we have
a total squared moment of $M^2=M_z^2+M_{dynamic}^2=16+10 =26~\mu_B^2$, which
is supposed to be $S(S+1)g^2\mu_B^2$ to satisfy the sum rule (see Methods). 
This
corresponds to $S\sim 2.1$ per Fe, in good agreement with the expected
$S=5/2$ from $Fe^{3+}$. On the other hand, if a simple Fe$^{3+}$ magnetic 
form factor were used, we would have obtained a $S_Z=1.4$ and $S=1.4$ instead,
which are much lower than the expected values.

Overall, our work demonstrates that the electromagnon excitations
in BiFeO$_3$ can be very well modeled with a simple spin-wave model
from a G-type AFM at low temperature. Despite the cycloid modulation
induced by coupling to the electric polarization, most  of the spin dynamics 
are barely modified, and consistent with the behaviour of a conventional
spin-wave. Upon heating decoherence of 
spin-wave excitations starts to occur even at just below room temperature (300~K).
At 580~K which is well below T$_N$, the coherence mode is already 
almost completely destroyed. In addition, a strong  hybridization
between the Fe and O orbitals needs to be taken into account when the magnetic
response is considered. Even as a good insulator, the electrons responsible 
for the magnetic moments are not restricted at the Fe sites, and 
the local magnetic moment could extend well beyond the typical ionic radius 
of Fe$^{3+}$ into the adjacent O$^{2-}$ sites.

\section{Methods}

Single crystal samples of BiFeO$_3$ are grown by floating zone technique at 
BNL. The crystal used in the measurement has a cylindrical shape, and  
weighs about 3.5~g, with a mosaic less than 2 degrees. 
Inelastic neutron scattering experiments are performed 
on the ARCS time-of-flight
spectrometer at the Spallation Neutron Source (SNS) at Oak Ridge National 
laboratory. Incident energies of 60~meV and 120~meV are used to probe the lower 
and higher part of the spin excitation.

The spin-wave calculations are based on a classical spin-wave model 
assuming a $S=5/2$ G-type AFM~\cite{Igor_Spinwave}. Based on the model, 
the values of 
$J_1$, $J_2$, and $J_3$ can all be directly derived from the data if the 
energies of the spin-wave excitations can be accurately determined at 
each wave-vector. For instance, if we define ${\bf Q_1=Q_{AF}}+(0.25,0.25,0)$,
${\bf Q_2=Q_{AF}}+(0.25,0.25,0.25)$,
${\bf Q_3=Q_{AF}}+(0.5,0.25,0)$, based on a simple spin-wave 
calculation, we have $J_1=\frac{1}{4S}(E_2^2-E_1^2)^{1/2}$,
$J_2=\frac{1}{8S}(E_2-E_3)$, and $J_3=\frac{1}{4}(\frac{E_3}{4S}-3J_1+8J_2)$.
Here $E_1$, $E_2$, and $E_3$ are spin-wave energies at ${\bf Q_1}$, ${\bf Q_2}$, 
and ${\bf Q_3}$, respectively. 

After normalizing the data using phonon 
intensities which have been obtained from the same measurements, a sum rule 
applies:
$\int\limits_{BZ}\int S({\bf Q},\hbar\omega)d^3{\bf Q}d(\hbar\omega)/
\int\limits_{BZ}d^3{\bf Q} = S(S+1)$, where $S({\bf Q},\hbar\omega)$
is the dynamic spin correlation function, 
and the integral is performed in one Brillouin zone.
In our case $I({\bf Q},\hbar\omega)$ = $\frac{2}{3}S({\bf Q},\hbar\omega)$,
where $I({\bf Q},\hbar\omega)$ is the normalized intensity.
The factor $\frac{2}{3}$ arises from assuming an isotropic dynamic spin 
correlation function (i.e. $S^{xx}=S^{yy}=S^{zz}$), and averaging the 
polarization factor of different domains. The scattering intensity can be 
modeled with the single mode approximation (SMA)~\cite{Igor_Neutron}, 
and the parameters 
that can vary include the exchange parameters $J_1$, $J_2$, $J_3$,
ground state binding energies for the nearest, second nearest (face diagonal),
and third nearest (body diagonal) spin pairs, $2J_1\langle S_0S_1\rangle$, 
$2J_2\langle S_0S_2\rangle$, $2J_3\langle S_0S_3\rangle$, and a finite 
energy broadening $\Gamma$ for high temperature data.

\bibliographystyle{naturemag}

\begin{thebibliography}{10}
\bibitem{LDA1}
\bibinfo{note}{We applied the WIEN2K implementation of the all-electron
  linearized augmented plane wave method in the LDA+U approximation using
  U=4~eV and J=0.9~eV. The basis size was determined by RKmax=7.0 and the
  Brillouin zone was sampled with a $7\times 7\times 7$ mesh.}
\end{thebibliography}

\end{document}